\documentclass[reprint]{revtex4-2}

\usepackage{amsmath}
\usepackage[colorlinks=true,linkcolor=blue,citecolor=blue,urlcolor=blue]{hyperref}

\begin{document}

\title{Quantum Covert Communication under Extreme Adversarial Control}

\author{Trey Li}
\email{trey.li@manchester.ac.uk}
\affiliation{University of Manchester, Manchester, UK}

\date{April 8, 2025} 

\begin{abstract} Secure quantum communication traditionally assumes that the adversary controls only the public channel. We consider a more powerful adversary who can demand private information of users. This type of adversary has been studied in public key cryptography in recent years, initiated by Persiano, Phan, and Yung at Eurocrypt 2022. We introduce a similar attacker to quantum communication, referring to it as the controller. The controller is a quantum computer that controls the entire communication infrastructure, including both classical and quantum channels. It can even ban classical public key cryptography and post-quantum public key cryptography, leaving only quantum cryptography and post-quantum symmetric key cryptography as the remaining options. We demonstrate how such a controller can control quantum communication and how users can achieve covert communication under its control.
\end{abstract}

\maketitle 

\section{Introduction} 

Quantum mechanics changes our understanding of information and security \cite{nielsen2010quantum}. Unlike classical security, which relies on computational complexity \cite{garey1979computers,sipser13}, quantum security is based on the fundamental laws of physics. The core principles of quantum mechanics—superposition, entanglement, uncertainty, wavefunction collapse, and no-cloning—form the foundation of secure quantum communication. 

Superposition \cite{wootters1982single} allows a photon to encode information in multiple quantum states simultaneously, similar to how a qubit can embody both 0 and 1 at the same time. This capability improves the efficiency and security of quantum information transmission by allowing for denser data encoding and more robust encryption methods. 

Entanglement \cite{einstein1935can,bennett1992communication} establishes a strong correlation between quantum particles. When one particle's state is measured, the state of its entangled partner is instantaneously determined, regardless of the distance between them. This principle enables secure remote agreement on photon states, making any unauthorized measurement immediately detectable.

Heisenberg's Uncertainty Principle \cite{heisenberg1927anschaulichen}  states that measuring certain conjugate properties (such as position and momentum, or energy and time) of a quantum system will inevitably disturb the system. This inherent disturbance exposes any interception or interference.

Wavefunction collapse \cite{zurek2003decoherence} happens when a quantum system is measured, forcing it into a single state and destroying the original superposition. This makes eavesdropping impossible, as the act of measurement collapses the state of the system. 

The no-cloning theorem \cite{wootters1982single,dieks1982communication} asserts that it is impossible to create an exact copy of an unknown quantum state. This safeguards data from duplication during transmission. 

Together, these principles form the foundation of quantum communication, providing security based on the fundamental laws of quantum mechanics rather than computation theory. 

However, security is never unconditional, it always depends on the assumed capacities of the adversary. 

Classical public key cryptography \cite{diffie2022new,rivest1978method} relies on computational problems like discrete logarithms \cite{diffie2022new} and factoring \cite{rivest1978method} . These problems are secure against classical adversaries but are vulnerable to quantum attacks \cite{shor1994algorithms,shor1999polynomial}.

Post-quantum cryptography \cite{bernstein2017post,chen2016report} uses computational problems that are believed to resist quantum attacks. While it aims to develop algorithms that remain secure in the quantum era, its implementation is classical, and it still relies on computational complexity without taking advantage of quantum mechanics. 

Quantum cryptography \cite{gisin2002quantum,pirandola2020advances,wehner2018quantum}, however, directly exploits quantum principles for secure communication. It resists both classical and quantum adversaries. Unlike classical and post-quantum cryptography, it does not depend on computational assumptions.

Early quantum protocols, such as BB84 \cite{bennett1984original} and E91 \cite{ekert1991quantum}, are designed to resists quantum eavesdroppers who can intercept, manipulate, and exploit entanglement. However, this model of adversary may not cover all future threats in quantum networks. 

A physically more powerful adversary was studied in \cite{bash2015quantum}. Unlike typical adversaries who might attempt to eavesdrop or tamper with the signal, this adversary exploits inherent channel imperfections to extract information. By leveraging channel noise, it can detect covert communication while passively monitoring transmissions without actively interfering. However, the strength of this adversary lies more in its physical capabilities rather than conceptual innovations. It remains an adversary that only focus on controlling the public channel without demanding private data from users.

In recent years, a conceptually super powerful adversary known as the dictator \cite{cryptoeprint:2022/639,2024_PERSIANO,cryptoeprint:2023/356} has been considered in public key cryptography. In contrast to traditional adversaries, this adversary can demand user private keys, aiming at a thorough monitoring of user communications. 

In this paper, we introduce it to quantum communication and refer to it as the controller in classical, quantum, and post-quantum cryptography. 

In public key cryptography, the controller is defined as a traditional adversary who further possesses users’ private keys \cite{cryptoeprint:2022/639,2024_PERSIANO}, violating the fundamental assumption of public key cryptography that user private keys are kept secret. 

In this paper, we consider an even more powerful controller who controls the entire communication infrastructure, including both classical and quantum channels. This controller can force users to submit any local secrets for which it has evidence of existence. Additionally, it is itself a quantum computer.

More specifically, anticipating future challenges, we envision an extremely powerful controller who is not satisfied with merely possessing users’ private keys in public key cryptography. This controller completely bans both classical and post-quantum public key cryptography and further seeks to control quantum communication. Such a controller is a plausible scenario within the next 30 years, particularly in specialized contexts like low-frequency and high-security environments. 

We analyze how this controller could undermine existing quantum protocols and how users could achieve covert communication under its control. 

We will call the users Alice and Bob, the eavesdropper Eve, and the controller Colin.

\section{Covert Communication in BB84} 

Unlike public key cryptography, there is no clear or general concepts of public or private keys in quantum communication. Hence the definition of the dictator \cite{cryptoeprint:2022/639} in public key cryptography does not naturally extend to quantum communication. 

For quantum communication, it is perhaps more appropriate to define the controller in terms of specific protocols or specific types of protocols. 

\subsection{Controller for BB84}

We take BB84 \cite{bennett1984original} as an example and see what the controller can do to control quantum communication. 

BB84 was the first quantum cryptography protocol ever invented. The main idea of BB84 is to agree on random photon states through random basis matching. It runs as follows:

\begin{enumerate}
\item {\bf Quantum Transmission Phase:} 
\begin{itemize}
\item Alice generates two random strings $a = (a_1,\dots,a_n) \in \{0,1\}^n$ (basis selection string) and $b = (b_1,\dots,b_n) \in \{0,1\}^n$ (state selection string), prepares a sequence of photons $|\varphi_{a_1b_1}\rangle,\dots, |\varphi_{a_nb_n}\rangle$ such that $|\varphi_{00}\rangle = |0\rangle$, $|\varphi_{01}\rangle = |1\rangle$, $|\varphi_{10}\rangle = |+\rangle$, and $|\varphi_{11}\rangle = |-\rangle$, and sends them to Bob over a quantum channel. 
\item Bob generates a random basis selection string $a' = (a_1',\dots,a_n') \in \{0,1\}^n$ and measures each photon $|\varphi_{a_ib_i}\rangle$ in the computation basis $\{|0\rangle,|1\rangle\}$ if $a_i' = 0$ or in the Hadamard basis $\{|+\rangle,|-\rangle\}$ if $a_i' = 1$, resulting in a state characteristic string $b' = (b_1',\dots,b_n')$ where $b_i' = 0$ if $|\varphi_{a_ib_i}\rangle = |0\rangle$ or $|+\rangle$, or $b_i' = 1$ if $|\varphi_{a_ib_i}\rangle = |1\rangle$ or $|-\rangle$.
\end{itemize}
\item {\bf Parameter Estimation Phase:} 
\begin{itemize}
\item Alice and Bob disclose their basis selection strings $a$ and $a'$ in a classical channel, discard the bits in $b$ and $b'$ where $a$ and $a'$ do not match, resulting in substrings $c \in \{0,1\}^k$ and $c'\in\{0,1\}^k$ of $b$ and $b'$. 
\item Alice randomly parses $c$ into substrings $d \in \{0,1\}^{k_1}$ and $e \in \{0,1\}^{k_2}$, where $k_1 < k_2$, and discloses the positions of $d$ to Bob.
\item Bob parses $c'$ in the same way to get the corresponding substrings $d'$ and $e'$ of $c'$. 
\item Alice and Bob disclose $d$ and $d'$ to calculate and estimate the error rate.  
\end{itemize} 
\item {\bf Key Reconciliation Phase:} 
\begin{itemize} 
\item If the error rate is not higher than a predefined threshold $t$, they use information reconciliation (e.g., an error correcting code) to turn $e$ and $e'$ into a shared secret $s$ and then use privacy amplification (e.g., a hash function) to turn $s$ into a shared key $K$; otherwise if the error rate is higher than $t$, they abandon this communication. 
\end{itemize}
\end{enumerate} 

Note that there are three main secrets in BB84:  

\begin{enumerate}
\item the final shared secret $s$;
\item the state selection string $b$;
\item the basis selection string $a$.
\end{enumerate}

We define three controllers to demand one of the three secrets respectively. Specifically:

\begin{enumerate}
\item []\textbf{Controller I:} The first controller demands $s$ from Alice and Bob after the BB84 execution is completed;
\item []\textbf{Controller II:} The second controller demands Alice's $b$ before she sends photons to Bob;
\item []\textbf{Controller III:} The third controller demands Alice's $a$ before she sends photons to Bob.
\end{enumerate} 

If the quantum channel is noise-free, each of the three controllers is progressively stronger than the one before. To see this, notice that $s$ can be computed from $b$ by monitoring the parameter estimation phase; and $b$ can be computed from $a$ by measuring Alice's photons using $a$. Note that in the latter case, to make sure that Alice and Bob do not abandon the communication, the controller can re-prepare photons according to $a,b$ and forward them to Bob.  

Let us see how the controller can do these in practice. First of all, the channels between the users and the controller must be quantum, otherwise the security between users boils down to classical security between the users and the controller. 

Therefore, the first controller can simply require Alice and Bob to use DL04 \cite{deng2004secure} (or any other quantum secure direct communication (QSDC) protocol) to pass $s$ to him; the second controller can require Alice to use BB84 (or any other QKD protocol) to agree with him on a shared secret and use this shared secret as $b$ in her BB84 communication with Bob; and the third controller is similar to the second except that Alice is required to use the shared secret as $a$ instead of $b$. 

We analyze if all the three controllers make sense. The first controller is the most practical one since it does not interfere the BB84 process. The second and third controllers have the risk of violating the purpose of using quantum communication, which is to avoid ``any'' trusted third party by the no-cloning theorem and wavefunction collapse. If the states or bases of the photons are disclosed to a third party in advance, the quantum advantage is entirely lost. Then quantum cryptography will be far less appealing, effectively reducing to classical cryptography. Then there will be no reason for the users to use quantum communication, as classical cryptography with simpler infrastructure would be a more preferable option.  

However, if we recall that the controllers will ban classical public key cryptography and post-quantum public key cryptography, leaving quantum cryptography and post-quantum symmetric key cryptography as the only options, then the two controllers are plausible. Thus it is highly valuable if we could find a solution to achieve covert communication under the control of such strong controllers. 

\subsection{Covert communication in BB84} 

Now we consider how to achieve covert communication against the first controller. We will handle the strongest third controller in DL04 in the next section.  

To handle the first controller, a naive idea might be for Alice and Bob to generate a fake shared secret using an input-length-preserving hash function. 

Specifically, suppose $s \in \{0,1\}^\ell$ is the shared secret between Alice and Bob after parameter estimation. Both of them hash $s$ into a fake shared secret $\bar  s = \bar  H(s)\in \{0,1\}^\ell$ of the same length, and submit $\bar  s$ to Colin. Now Alice and Bob have two shared keys: $K = H(s)$ and $\bar  K = H(\bar  s)$, where $H$ is the hash function used in normal BB84. Colin knows the fake shared key $\bar K$ but he does not know the true shared key $K$. 

However, this approach does not work because the controller could verify the shared key by inspecting ciphertexts in subsequent communications between Alice and Bob. This would leave no opportunity for Alice and Bob to use their true shared key $K$. 

Our real method is to embed message into the error checking phase. The high level idea is that after photon transmission and bases publishing, Alice assumes that the channel is noise-free and immediately generates a shared key and uses it to encrypt messages and publishes the ciphertext for error rate estimation; on the other side, Bob simultaneously publishes a random string for error rate estimation, and he can decrypt Alice's ciphertext using the shared key. 

The detailed scheme is as follows, where we modified the error checking phase and key generation phase of BB84. 

\begin{enumerate}
\item {\bf Quantum Transmission Phase:} 
\begin{itemize}
\item Same as standard BB84.
\end{itemize} 
\item {\bf Parameter Estimation Phase:} 
\begin{itemize}
\item Alice and Bob disclose their basis selection strings $a$ and $a'$ in a classical channel, discard the bits in $b$ and $b'$ where $a$ and $a'$ do not match, resulting in substrings $c \in \{0,1\}^k$ and $c'\in\{0,1\}^k$ of $b$ and $b'$. 
\item Alice randomly parses $c$ into substrings $d \in \{0,1\}^{k_1}$ and $e \in \{0,1\}^{k_2}$, where $k_1 < k_2$, uses error correcting code to correct $e$ to get $f \in \{0,1\}^\ell$ with $k_1< \ell < k_2$, hashes it to get $h = H(f) \in \{0,1\}^{k_1}$, encrypts message $m \in\{0,1\}^{k_1}$ as $ct = h \otimes m$, where $\otimes$ denotes bit-wise XOR. Simultaneously, Bob samples a random string $r \in \{0,1\}^{k_1}$.
\item Alice and Bob publish $ct$ and $r$ for error rate estimation.
\item Bob parses $c'$ into $d' \in \{0,1\}^{k_1}$ and $e' \in \{0,1\}^{k_2}$, where $k_1 < k_2$, uses error correcting code to correct $e'$ to get $f'\in \{0,1\}^\ell$ with $k_1< \ell < k_2$, hashes it to get $h' = H(f') \in \{0,1\}^{k_1}$, and tries to decrypt $ct$ by computing $m' = ct \otimes h'$.
\end{itemize} 
\item {\bf Key Generation Phase:} 
\begin{itemize}  
\item If the error rate between $ct$ and $r$ is not higher than a predefined threshold $t$, Alice hashes $f$ into $\bar  f = \bar  H(f) \in \{0,1\}^\ell$ and submits it to the controller, simultaneously, Bob hashes $f'$ into $\bar  f' = \bar  H(f') \in \{0,1\}^\ell$ and submits it to the controller; otherwise if the error rate is higher than $t$, they abandon the communication and submit random strings to the controller. 
\end{itemize} 
\end{enumerate} 

The error check will fail with high probability because $ct$ and $r$ are independently random strings, and the expected number of matching bits is $k_1/2$. For instance, using similar parameters as in \cite{yuan2005continuous}, with $k_1 = 5000$ and $t = 10\%$, the probability that the number of unmatched bits between $ct$ and $r$ is smaller than 500 is approximately $10^{-100}\approx 2^{-332.19}$, a rare event. 

The controller cannot distinguish whether the failure is caused by covert communication or eavesdropping. This is because, by the no-cloning theorem and wavefunction collapse, both cases result in a random $r$. Also, by the same reason, Colin has no idea whether the one-time pad ciphertext $ct$ is truly a valid substring of $b$, unless he had used Alice's secret basis (which he does not know) to eavesdrop Alice's photons. 

If the error checking accidentally passes, Alice and Bob submit fake shared keys $\bar  K$ to Colin in a way similar to the naive approach. Again, by no-cloning and wavefunction collapse, Colin has no idea whether $\bar K$ is the true shared secret. 

Overall, in a one-time execution of the revised BB84, Colin statistically / informationally has no advantage in telling whether Alice and Bob are exchanging covert information. 

As to the security against a typical attacker, note that Colin can be viewed as an upgraded Eve, with access to additional information beyond Eve can. Hence if Colin cannot detect whether Alice and Bob are exchanging covert messages, then Eve cannot gain any information of the covert message either. 

Similar to traditional attacks in quantum communication, both Colin and Eve can disrupt the communication by eavesdropping on the photons. But they cannot learn anything about the covert messages. 

\section{Covert key distribution in DL04} 

We consider the third controller for DL04 \cite{deng2004secure}, this controller knows the secret basis of the photon sender. 

\subsection{Controller for DL04}

DL04 is essentially BB84 with photon encoding and returning. Its main idea is to encode each message bit into a photon by flipping its state using unitary operations.  It runs as follows: 

\begin{enumerate}
\item {\bf Photons Sending Phase:} 
\begin{itemize}
\item Bob generates two random strings $a = (a_1,\dots,a_n) \in \{0,1\}^n$ and $b = (b_1,\dots,b_n) \in \{0,1\}^n$, prepares a sequence of photons $p = (|\varphi_{a_1b_1}\rangle$,..., $|\varphi_{a_nb_n}\rangle)$ such that $|\varphi_{00}\rangle = |0\rangle$, $|\varphi_{01}\rangle = |1\rangle$, $|\varphi_{10}\rangle = |+\rangle$, and $|\varphi_{11}\rangle = |-\rangle$, and sends them to Alice over a quantum channel. Call this batch of photons the A-batch. 
\item Alice randomly chooses a batch of photons to complete a typical BB84 error rate estimation process with Bob. Call this batch of photons the S-batch. 
\item If the error rate is too high, Bob concludes that the communication is insecure, and the protocol is aborted. If the error rate is sufficiently low, the remaining photons, called the B-batch (suppose its size is $k$), will be used to do encryption. 
\end{itemize}

\item {\bf Photons Returning Phase:} 
\begin{itemize}
\item After confirming the security of the B-batch, Alice proceeds to encode the message $m = \{0,1\}^k$ by encoding each photon in the B-batch with one of two possible operations:
\begin{itemize}
\item [$\bullet$] $ I = \lvert 0 \rangle \langle 0 \rvert + \lvert 1 \rangle \langle 1 \rvert $
\item [$\bullet$] $ U = i \sigma_y = \lvert 0 \rangle \langle 1 \rvert - \lvert 1 \rangle \langle 0 \rvert $
\end{itemize}
where if $m_i = 0$ then apply $I$ on the $i$-th photon of the B-batch, otherwise apply $U$ on it, resulting in a batch of encoded photons. 
\begin{itemize}
\item The operation $I$ does not have any affects, and $U$ has the following effects on the states: \begin{itemize}
\item [$\bullet$] $ U |0\rangle = -|1\rangle, \quad U |1\rangle = |0\rangle $
\item [$\bullet$] $ U |+\rangle = |-\rangle, \quad U |-\rangle = -|+\rangle $
\end{itemize} 
\end{itemize} Alice sends the encoded B-batch to Bob. 
\item Bob measures the photons using the correct bases which he knows, resulting in a string $d = (d_1,\dots,d_k)\in \{0,1\}^k$. Bob then extracts the subsequence $e \in \{0,1\}^k$ of $a$ corresponding to the positions of the B-batch in the A-batch, and recovers the message as $m' = d \otimes e$.
\item To ensure the security of the transmission, Alice randomly selects some photons from the B-batch as checking instances to check the error rate with Bob. 
\end{itemize} 
\end{enumerate} 

There are also three main secrets in DL04: 

\begin{enumerate}
\item the message $m$; 
\item the state selection string $b$; 
\item the basis selection string $a$. 
\end{enumerate} 

We consider a very strong controller who demands all three of them (in principle, he can demand everything he has evidence of existence). Specifically:

\begin{enumerate}
\item [] {\bf Controller IV:} This controller requires Bob to share two secrets $a,b$ with him using BB84 before starting DL04 with Alice, and requires him to use $a,b$ as the basis selection string and state selection string. He can then use $a$ to measure Bob's sending photons and see if their states match $b$; he then re-prepares the same photons according to $a,b$ and forward them to Alice. He can also require Alice to send him the message $m$ she wants to send to Bob anytime before she returns Bob's photons, and measures her returning photons to check if she is really sending $m$ to Bob; he then re-prepares the same B-batch and re-encodes the message into them and forward them to Bob. 
\end{enumerate} 

As we can see, everything in the standard DL04 protocol is basically transparent to this controller. 

\subsection{Covert Key Distribution in DL04}

We use entangled photons and Bell states to achieve covert key distribution under the control of this Colin. The idea is to carry a random covert message $m'$ (the key that Alice wants to share with Bob) by the Bell states of entangled photons. More specifically, we let the A-batch to be the signal photons of a set of entangled photons of random Bell states. Alice encodes an extra bit into each photon in the B-batch by flipping the bit different times, which will result in a change of the Bell state of the entangled pair, and only Bob, who has the idler photons can measure the Bell states of the entangled pairs, and thus tell any Bell state changes. Colin has no idea about the Bell state of the entangled pairs since he does not have the idler photons. He will have zero advantage in telling any Bell state changes. The revised DL04 is as follows. 

\begin{enumerate}
\item {\bf Photons Sending Phase:} 
\begin{itemize} 
\item Bob generates three random strings: 
\begin{itemize}
\item $a = (a_1, \dots, a_n) \in \{0,1\}^n$, the basis selection string.
\item $b = (b_1, \dots, b_n) \in \{0,1\}^n$, the state selection string.
\item $c = (c_1, \dots, c_n) \in \{0,1\}^n$, the Bell state selection string.
\end{itemize}
\item Bob prepares $n$ entangled photon pairs $(A_i, B_i)$ in the Bell state $ |\Phi^{c_i}\rangle_{A_i B_i}$ (if $a_i = 0$) or $|\Psi^{c_i}\rangle_{A_i B_i}$ (if $a_i = 1$), where: 
$$
|\Phi^{c_i}\rangle_{A_i B_i} = \frac{1}{\sqrt{2}} \left( |0\rangle_{A_i} |0\rangle_{B_i} + (-1)^{c_i} |1\rangle_{A_i} |1\rangle_{B_i} \right)
$$
$$
|\Psi^{c_i}\rangle_{A_i B_i} = \frac{1}{\sqrt{2}} \left( |0\rangle_{A_i} |1\rangle_{B_i} + (-1)^{c_i} |1\rangle_{A_i} |0\rangle_{B_i} \right)
$$
\item Bob selects signal photons $A = (A_1, A_2, \dots, A_n)$ according to $b$ and sends them to Alice, where the selection is as follows:
\begin{itemize}
\item If $b_i = 0$, the photon that will be measured as $|0\rangle$ (if $a_i = 0$) or $|+\rangle$ (if $a_i = 1$) is marked as the signal photon $A_i$ and the other is marked as the idler photon $B_i$.
\item If $b_i = 1$, the photon that will be measured as $|1\rangle$ (if $a_i = 0$) or $|-\rangle$ (if $a_i = 1$) is marked as the signal photon $A_i$ and the other is marked as the idler photon $B_i$.
\end{itemize}
\item Alice and Bob performs an error rate estimation process as in the standard DL04 using an S-batch. 
\end{itemize} 

\item {\bf Photons Returning Phase:} 
\begin{itemize}
\item After confirming the security of the B-batch, Alice proceeds to encode the normal message $m = \{0,1\}^k$ and random covert message $m' = \{0,1\}^k$ by flipping the photons in B-batch with the unitary operation:
\begin{itemize} 
\item [$\bullet$] $U = i\sigma_y = \lvert 0 \rangle \langle 1 \rvert - \lvert 1 \rangle \langle 0 \rvert$
\end{itemize} for $0,1,2,3$ times depending on whether $m_im_i'$ is $00$, $01$, $10$ or $11$.

\begin{itemize}
\item Applying $U$ on $|0\rangle$ and $|+\rangle$ for $0,1,2,3$ times results in single photon state/phase changed as:        \begin{itemize}
\item [$\bullet$] $ |0\rangle \xrightarrow{U} -|1\rangle \xrightarrow{U} -|0\rangle \xrightarrow{U} |1\rangle$
\item [$\bullet$] $ |+\rangle \xrightarrow{U} |-\rangle \xrightarrow{U} -|+\rangle \xrightarrow{U} -|-\rangle$
\end{itemize} 
It results in Bell state changed as:
\begin{itemize} 
\item [$\bullet$] $|\Phi^+\rangle \xrightarrow{U} |\Phi^-\rangle \xrightarrow{U} -|\Phi^+\rangle \xrightarrow{U} -|\Phi^-\rangle$
\item [$\bullet$]$ |\Psi^+\rangle \xrightarrow{U} |\Psi^-\rangle \xrightarrow{U} -|\Psi^+\rangle \xrightarrow{U} -|\Psi^-\rangle$
\end{itemize}
\end{itemize} 

\item Alice then sends the encoded B-batch, $A' = (A'_1, A'_2, \dots, A'_k)$, back to Bob.
\end{itemize}

\item {\bf Decoding Phase:} 
\begin{itemize}
\item Bob can recover one of $m$ and $m'$ but not both. Bob can recover the normal message bit $m_i$ by checking if the state of the signal photon $A_i'$ has flipped. Bob can recover the covert message bit $m'_i$ by checking if the Bell state of the entangle pair $(A_i',B_i)$ has flipped. 
\item If Bob chooses to recover $m$, he performs the second error rate estimation with Alice normally as standard DL04. If Bob chooses to decode $m'$, he performs the second error rate estimation using a random string (and they will announce failure with high probability). 
\end{itemize} 
\end{enumerate}

The correctness of the protocol can be seen from the explanation embedded in the protocol description. 

Let us analyze what Colin can do to detect the existence of covert communication, and show why the protocol is secure (i.e., covert communication undetectable). 

Colin can measure Bob's photon on their way to Alice and see if they are single photons (no entangled photons are expected in normal DL04). However, without the idler photons, Colin cannot distinguish whether the signal photons are single or entangled. 

Colin may also want to check if the photons from Bob to Alice have any unexpected phases (no ``$-$'' phases are expected in normal DL04). But this does not give him any evidence that Bob and Alice are trying to do covert communication, since the phase changes can also be caused by Eve, and the two causes will result in the same distributions, which are indistinguishable. 

Colin can also clean the sending photons by preparing new photons according to $a$ and forward them to Alice. But this only disrupts potential covert communication and not solidifying it. Disruption is inevitable in any quantum protocol anyways. 

Colin can also measure Alice's returning photons on their way to Bob and see if there are any unexpected phases. Again, this does not give any evidence to Colin, since only Bob knows the initial phases of the photons, and without this information, Colin can only see random phase changes, and he cannot tell whether this is due to covert communication or phase interference by Eve. 

A more powerful detect would be for Colin to measure Bob's sending photons, recover $b$, re-prepare new entangled photons according to strings $a,b$ as well as a newly sampled random Bell state selection string $c'$, and forward a new A-batch to Alice. Bob will know that his sending photons are being intercepted, but he cannot signal Alice in any way, particularly during the first error rate estimation phase, since this will alert Colin. Alice does not know the signal photons have been replaced, and she will encode both $m$ and $m'$ as scheduled, and returns the encoded B-batch to Bob. Colin then measures Alice's encoded photons and recovers her covert message $m'$ by measuring Bell state changes. He can do this because he has the corresponding idler photons.  

This detection does not work either. This is because $m'$ is a random message, Colin cannot tell if the Bell state changes are caused by covert communication or Eve. This is why this covert channel can only be used to distribute random keys rather than arbitrary messages. Otherwise if $m'$ is a low entropy message, Colin can immediately tell the existence of covert communication. 

In this detection, there is a more subtle way for Colin to detect covert communication. Since he has the idler photons, whenever Alice applies a unit operation on the signal photon, it will affect the idler photon. Colin can simply observe whether the same photon is being flipped multiple times within a very short time (in standard DL04 each photon is flipped by Alice one time at most). But this detection can be avoided if Alice chooses random timings to flip the signal photons in order to mimic eavesdropping from Eve. 

Note also that in this attack, Colin cannot eventually obtain a common string $m'$ between Alice and Bob, because he has destroyed the entanglement of Bob's initial pairs, there is no way he can fake a new batch of returning photons such that Bob can still recover $m'$, even if he knows all the secrets $a,b,m'$. 

This means that it is not only true that Colin cannot detect covert communication, but also true that Colin cannot obtain any secret key $m'$ shared by Alice and Bob's via this covert channel. He can only disrupt potential covert key sharing, which is what Eve can do as well in any quantum protocols via eavesdropping. 

Note that the other cover communication idea does not work, namely Bob uses another random basis selection string $a'$ different from the $a$ that he agreed with Colin beforehand. This is because in the first error rate estimation phase, Bob has to publish the $a$ that Colin expects, and for this $a$, Alice's measurement will not match Bob's true states with high probability. This means that the security check will not pass with high probability, and there will be no reason for Alice to continue the protocol and return any photons to Bob. 

Note also that the proposed covert communication idea cannot be directly applied in BB84. This is because with one pass of photons, even if Alice (in BB84, Alice is the photon sender) initially uses entangled photons and embeds Bell state changes into the entangled pairs, Bob will not have any advantage in recovering the message $m'$ over Colin or Eve. Bob's advantage only occurs after agreeing with Alice on some common secret after the BB84 error rate estimation phase. But that common secret is not known before Alice encodes the covert message $m'$. I.e., the common secret is independent of the encoding of $m'$ and thus useless in recovering $m'$. 

\section{Subsequent Covert communication}

In the previous two sections, we have shown how to perform covert quantum key distribution (where in BB84 we can send any covert messages, including random keys). 

We show in this section that, after an initial secret sharing, Alice and Bob can easily share more and more random secrets in subsequent communications. The idea is as follows. 

Suppose Alice and Bob has shared a random secret $m'$, resulting in a final shared key $K$ (e.g., via a hash function). Alice can then use this $K$ as the basis selection string $a$ in her next execution of BB84 (or any other quantum protocols) with Bob, and sends photons with random states. Denote the state selection string of these random states as $b$. Since Bob knows $a$, he can always measure the photons in the correct basis, and recover the state selection string $b' \approx b$ up to differences caused by noise. Assume the noise rate of the quantum channel is low, they can then use error correcting code to turn $b$ and $b'$ into a common secret $s$. This is the new secret they share. 

In order to fake the controller, in the error rate estimation phase, Alice will publish the basis that Colin forces her to use, and Bob can simply publish a random basis. They then (with high probability) announce security check failure.  

They can repeat this method for arbitrary many times. Thanks to the no-cloning theorem, leaking one of the secrets will not expose any other historical shared secrets because no one can keep a record of the historical photons. 

We may have realized that this method can be used to transmit any message $b$ rather than random $b$; and this method generally works in any quantum protocol that involves photon transmission. 

If Alice wants to send an arbitrary message $m$, to further secure the communication, Alice can use a post-quantum symmetric key encryption scheme (e.g. AES \cite{rijmen2001advanced}) to encrypt the message into a psuedorandom ciphertext $b$, then use $b$ as the state section string. 

\section{Conclusive General Solution} 

We have shown quantum cover communication in concrete protocols. Note that these methods are not limited to the discussed protocols, they can be applied to a broad range of protocols, providing a general solution to quantum covert communication. 

Specifically, the method on BB84 can be applied to all quantum communication protocols that involve a (public) parameter estimation phase similar to BB84. For example, E91 \cite{ekert1991quantum}, DL04 \cite{deng2004secure}, and many more. The method on DL04 can be applied to almost all quantum communication protocols that involve photon sending and returning. The method for subsequent covert communication can be applied to almost all quantum communication protocols that involve photon sending. 

Now we summarize the methods and conclude a general solution to quantum cover communication that works for most (if not all) quantum protocols that involve photons sending and/or returning. 

The controller is defined as:

\begin{enumerate}
\item [] {\bf Controller V:} It is a quantum computer that bans classical public key cryptography and post-quantum public key cryptography, and controls quantum cryptography by demanding three secret strings from the photon sender: the basis selection string $a$, the state selection string $b$, and the Bell state selection string $c$ (if the protocol originally uses entangled photons). 
\end{enumerate} 

The solution to quantum covert communication under this controller includes three stages:  

\begin{enumerate}
\item {\bf Quantum Covert Key Distribution}: Using entangled photons similar to the revised DL04 (regardless of whether the protocol originally uses entangled photons).
\item {\bf Subsequent Quantum Covert Key Distribution}: Setting the state selection string to be the new secret intended to distribute, and the basis selection string to be the common key shared via stage 1.
\item {\bf Quantum Covert Communication}: Setting the state selection string to be the message itself or the ciphertext of the message encrypted by a post-quantum symmetric key encryption scheme using a key shared via stage 1 or 2, and the basis selection string to be (another) key shared via stage 1 or 2.
\end{enumerate} 

As we analyzed earlier, with the use of entangled photons, any third party, including Colin, will not be able to detect covert communication, even if he requires the sender to use pre-agreed strings $a$, $b$, and $c$. The photon sender can still use different strings $a'$, $b'$, and $c'$ to achieve covert communication and attribute all unusual changes to Eve. Unless Colin can control any Eve or predict the behaviors of any Eve, he is not expected to have any advantage in solidifying the existence of covert communication. This completes our discussion of quantum covert communication.

\end{document}